


\input harvmac
\def\half{{1 \over 2}}

\def\bt{$\bullet$}

\Title{HUTP-92/A052}{\vbox{\centerline{HOW THE LITTLE GROUP CONSTRAINS}
\vskip2pt\centerline{MASSLESS FIELD REPRESENTATIONS}
\vskip2pt\centerline{IN HIGHER EVEN DIMENSIONS}}}

\centerline{{\bf Vineer Bhansali} \footnote{$^\dagger$}{Talk Presented at the
\it XIX International Colloquium on Group
Theoretical Methods in Physics, \rm Salamanca, Spain (June 1992).  Research
supported in part by
NSF Contract No. PHY-87-14654 and TNRLC Grant No. RGFY9206.}}
\bigskip\centerline{Lyman Laboratory of Physics}
\centerline{Harvard University}\centerline{Cambridge, MA 02138}

\vskip .2in
\centerline{\bf Abstract}
\medskip
\noindent
Assuming trivial action of Euclidean translations of the little group, we
derive a correspondence between massless field representations transforming
under the
full generalized even dimensional Lorentz group, and highest weight states of
the relevant little group. This yields a connection between `helicity' and
`chirality' in all dimensions, and highlights the special nature of the
restricted representations under `gauge' transformations.

\newsec{Introduction}

\noindent
The generalized Lorentz group $SO(2n+1,1)$
commutation relations in $d=2n+2$ spacetime dimensions are
\eqn\lorg{[J_{AB},J_{CD}]= i(\delta_{AC}J_{BD} + \delta_{BD}J_{AC} -
\delta_{AD}J_{BC}-\delta_{BC}J_{AD} )}
with $A,B=0, \ldots (d-1)$, $\delta_{AB} = 1 \, {\rm if} \, A = B; \, 0 \, {\rm
otherwise} $; ${K_i=iJ_{i0}= - iJ_{0i}} $ are boosts in the $i$th direction,
and for $i,j \neq 0$, $J_{ij} = - J_{ji}$ is
the rotation generator in the $i,j$
hyperplane.
When required, to `Wick' rotate the noncompact
algebra to the compact algebra of the special orthogonal algebra $SO(2n+2)$,
or vice versa, make the replacements
$SO(2n+1,1) \leftrightarrow SO(2n+2)$,
$i J_{i0} \leftrightarrow J_{i,d}$ and $g  \leftrightarrow \delta $,
where $g = {\rm diag} \, (-1,1,\ldots,1,1)$.
The compact form
$SO(2n+2) \equiv D_{n+1}$ has
$ {\rm rank} \ = \ n+1$,
$ {\rm dimension} \ = \ (n+1) (2n+1)$,
${\rm \# \, of \, roots} \ = \ 2 n (n+1)$,
${\rm \# \, of \, positive \, roots} \ = \ n(n+1)$, and
${\rm \# \, of \, simple \, positive \, roots} \ = \ (n+1) $, and the
members of the last set are the linearly
independent raising operators.
\noindent
For a massless particle in $d$ dimensions moving in the $d-1$ direction,
the relevant \it little group \rm of Wigner is defined to be the maximal
subgroup of the Lorentz group that leaves invariant its `standard' momentum
$k_{\mu} = (k, 0, \ldots , k); \mu=0,\ldots, d-1$. The little group is
generated by the $n$ commuting rotations $J_{12},
J_{34}, \ldots, J_{(2n-1,2n)}$ which will be the Cartan
generators (and hereafter the
`weight notation', denoted by a subscript $W$, will
refer to the specification of a state in terms of the
eigenvalues of this ordered set of Cartan generators), \it and \rm the $2n$
`translation' generators
\eqn\trans{L^+_i \equiv J_{i,2n+1} + i J_{i,0}, \, i = 1,\ldots 2n}
which, using \lorg\ can be seen to form an Abelian subalgebra
\eqn\as{[L^+_i,L^+_j] = 0.}
Each translation indexed by $i$ is a sum of a boost in the
$i$th direction and a rotation in the $i, d-1$ plane.
The commutation relations of the little group are
\eqn\lgcr{\eqalign{[L^+_i , L^+_j] &= 0 \cr
[J_{ij}, L^+_k] &= i ( \delta_{ik} L^+_j - \delta_{jk} L^+_i) \cr
[J_{ij}, J_{kl}] &= i (\delta_{ik} J_{jl} - {\rm permutations}) .\cr}}
The little group \lgcr\ is not semi-simple (it has
an Abelian subalgebra of translations) and is isomorphic to
 $E(d-2)$, the Euclidean group
in $d-2$ dimensions.  The generators
\eqn\trami{L^-_i \equiv J_{i,2n+1} - i J_{i,0} =L^+_i - 2 i J_{i,0} }
 also form an Abelian
subalgebra, since
\eqn\tramia{[L^-_i,L^-_j] = 0,}
but they do not belong to the little group for the choice of
standard momentum we have made. Note also the useful facts that
\eqn\lt{\eqalign{[L^+_i,L^-_j] &= 2 i J_{ij}, i\neq j \cr
&=2 J_{2n+1,0}, i = j  \cr}}
and under complex-conjugation
\eqn\cce{\eqalign{(L^+_i)^{\ast} &= - L^-_i \cr
(L^-_i)^{\ast} &= - L^+_i .\cr }}
\noindent
We now state a key assumption: \bf the translation generators annihilate
physical states. \rm  There are a number of reasons for this:
(1) \it Finite dimensional representations only: \rm
The group transformation
corresponding to
the translation $L^+_i$, is written as $D^+_i(\chi_i) = e^{-i \chi_i L^+_i}$;
the dimensionality of the representation
 is characterized by the length of the translation vector $\sqrt{\sum_i
\chi_i^2}$. For finite dimensional representations the translation parameter
$\chi_i = 0, \forall i$.
(2) \it Gauge-independence: \rm
In general, an eigenstate of the Cartan generators of
the little
group $J_{12}, \ldots J_{(2n-1, 2n)}$ is not an eigenstate of the translation
generators $L^+_i$ since they do
not mutually commute.
The eigenstates of the Cartan generators can be written as the $d$ dimensional
polarization vectors (e.g. in four dimensions the generator $J_{12}$ which
generates $z$-rotations has the eigenvectors $\epsilon^{\mu} = (0,1,\pm
i,0)$). It can be checked then, for instance, that the transversality condition
required for the photon vector potential to be a Cartan eigenstate is not
invariant under
finite translations. In fact, the translations generate effects
identical
to Abelian `gauge'
transformations \ref\WEINB{S. Weinberg, Phys. Rev. 135, B1049
(1964)} \ref\HKS{D. Han, Y.S. Kim, Am. J.
Phys. 49(4), 348(1981); D. Han, Y.S. Kim, D. Son, Phys. Rev. D25, 461 (1982);
D.
Han, Y.S. Kim, D. Son, Phys. Rev. D31, 328 (1981); D.Han, Y.S. Kim, D. Son,
Phys. Rev. D26, 3717 (1982)} .
Thus the requirement of trivial translations is the
requirement
that only `gauge independent' objects be considered.
These two consequences of trivial translations are
physically desirable from the point of view of making a scattering theory for a
finite number of physical degrees of freedom without auxiliary conditions, as
in
\ref\CG{S. Coleman, B. Grossman,
Nucl. Phys. B203, 205 (1982)} \ref\VB{V. Bhansali, Mod. Phys. Lett. A5, 1223
(1990)}. (3) \it `Factorization' of invariant operators: \rm
It can also be shown as a rather nontrivial consequence \ref\vbf{V. Bhansali,
in
\it Classical and Quantum Systems: Foundations and Symmeteries \rm Proceedings
of Second International Wigner Symposium, Goslar, Germany (1991), World
Scientific. } that the eigenvalue of an invariant operator in the enveloping
algebra of the higher dimensional Lorentz group factorizes into the eigenvalue
of a generalized Pauli-Lubanski pseudovector and and a simple factor related
to the boost generator.

\newsec{The Main Theorem}

\noindent
Only the main results are
highlighted here and the interested reader is referred for further details
to \ref\vbj{V. Bhansali, Intl. Jour. Mod.
Phys. A7(26) (1992)}.
Define a \it physical field \rm $\Lambda$
as a representation transforming under the
 full higher dimensional Lorentz group
and obeying the condition of trivial translations
${L^+_i \Lambda = 0, \, i=1,2 \ldots 2n}$.
\smallskip\noindent
\bf Main Theorem: \it   A physical field $\Lambda$ is a highest weight of
the Lorentz group.
\smallskip
\noindent
\bf Proof: \rm Take the full group to be $SO(2n+1,1)$.  The little group is
then $E(2n)$.  By definition, a physical field is annihilated by all the
translations $L^+_1, \ldots L^+_{2n}$.  Now, a highest weight is by
definition the state annihilated by \it all \rm the linearly independent
raising operators.  For $SO(2n+1,1)$,
which is rank $n+1$, we need thus to find the $n+1$ positive simple roots and
show that they all annihilate the physical field. To this end, we first want
to prove the following assertion:
\smallskip
\noindent
\bf Lemma: \it All the raising operators can be made using linear
combinations of the translation generators. \rm
\smallskip
\noindent
\bf Proof of Lemma: \rm  Since there are $2n$ available translations, and
$n+1$ required linearly independent raising operators, for $n \geq 1$, i.e.
for four dimensions or more, there are certainly enough translations available
to make all the linearly independent raising operators.
We choose the Cartan generators in $SO(2n+1,1)$
to be $J_{12}, \ldots , J_{2n-3,2n-2}, J_{2n+1,0}$. We claim that the
 complete set of
linearly independent raising operators for $n>1$ is
\eqn\ras{\eqalign{&L^+_1 \pm i L^+_2 \cr
 &L^+_3 + i L^+_4 \cr
&\vdots\cr
&L^+_{2n-1} + i L^+_{2n} .\cr}}
For example, in four dimensions ($n=1$) the two linearly
independent
raising operators are $L^+_1 \pm i L^+_2$.  To check, in any dimension
 that \ras\ is
indeed the complete set of linearly independent raising operators, commute
each member of the set
with the Cartan generators using \lorg\ and \lgcr\ to obtain the coordinates
of the positive simple roots in the Cartan basis.
Translating this to the Dynkin basis \vbj\ \ref\SL{R. Slansky, Phys. Rep. 79,
1 (1981)} obtain the rows of the Cartan
matrix.  But since the rows of the Cartan matrix are defined to be the
coordinates of the linearly independent raising operators under commutation
with the Cartan generators, \ras\ is in fact all of them.
With the physical field condition and \ras\ the theorem is proved. {\bf QED}
\smallskip\noindent
Note that using \lt\ we obtain
\eqn\lowe{[L^+_i + i L^+_{i+1}, L^-_i - i L^-_{i+1}] = 4 (J_{2n+1,0} +
J_{i,i+1}), \, i=1,3,5,2n-1}
and
\eqn\lowem{[L^+_1 - i L^+_2, L^-_1 + i L^-_2] = 4 (J_{12} - J_{30}).}
Since the right hand side lies in the Cartan subalgebra,
the lowering operator corresponding to each of the raising operators  is
obtained by replacing $L^+_i \rightarrow L^-_i, \, i \rightarrow -i$ in \ras.
With the definition
\eqn\spir{\eqalign{E^+_{1,1,0,\ldots,0} = A^+_+ &\equiv L^+_1 + iL^+_2 \cr
E^+_{-1,1,0,\ldots,0} = A^+_- &\equiv L^+_1 - iL^+_2 \cr
E^-_{1,-1,0,\ldots,0} = A^-_- &\equiv L^-_1 - iL^-_2 \cr
E^-_{-1,-1,0,\ldots,0} = A^-_+ &\equiv L^-_1 + i L^-_2 \cr}}
where the subscripts on the roots $E$ denote the eigenvalue under commutation
with the Cartan generators,
\eqn\ort{[A^+_+, A^+_-] = [A^+_+, A^-_+] = [A^-_+, A^-_-] = [A^+_-, A^-_-] =
0}
we note that $A^+_+$ and $A^+_-$ are raising operators in two orthogonal
directions (with $A^-_-$ and $A^-_+$ the orthogonal lowering operators).
 In four dimensions, $A^+_+, A^+_-$ are the two linearly independent
raising operators, and $A^-_-, A^-_+$ are the corresponding lowering
operators.
{}From the physical field condition
\eqn\fou{L^+_1 \Lambda = L^+_2 \Lambda = 0 \Leftrightarrow A^+_+ \Lambda =
A^+_- \Lambda = 0}
which shows that the physical field is a highest weight; complex
conjugating the last two equations with the use of \cce\ :
\eqn\lowest{A^-_- \Lambda^{\ast} = A^-_+ \Lambda^{\ast} = 0}
shows that the complex conjugate is the lowest weight. For
instance, the weight diagram for the left and right handed spinors in four
dimensions (in Dynkin notation) %
\eqn\leftspinor{\eqalign{{\rm Left} \ \ \ \ & {\rm Right} \cr
(+1 \ \ 0 ) \ \ \ \ & (0 \ \ +1) \cr
(-1 \ \ 0) \ \ \ \ & (0 \ \ -1) \cr }}
confirm the result that indeed the spinors are inequivalent and
self-conjugate
(i.e. the weight diagrams reflect to minus themselves).

\newsec{Helicity-Chirality Correlation in Higher Even Dimensions}

\noindent
In one of his classic papers on `Feynman Rules for Any Spin', Weinberg
\ref\WEIN{S. Weinberg, Phys. Rev. 134, B882 (1964)} proves that the
annihilation operator for a massless particle of helicity $\lambda$ and the
creation operator for the antiparticle  with helicity $-\lambda$ can only be
used to form a field transforming as
${U[\Lambda] \psi_n(x) U[\Lambda]^{-1} = \sum_m \
D_{nm}[\Lambda^{-1}] \psi_m(\Lambda x)}$
under those representations $[A,B]$ of $SO(3,1)$ such that $\lambda =
B-A$\foot{$A$ and $B$ correspond to independent $SU(2)'s$.}.  This restriction
arises solely due to the non-semi-simplicity of the little group for massless
particles; in particular due to the requirement that the Euclidean
`translations' of the little group act trivially on the physical Hilbert
space.  As a direct consequence of Weinberg's result it is observed that in
four-dimensions a physical left-handed, helicity $-j$ particle can only
correspond to a representation $[j+n,n]$, ($ n $ is an integral multiple of
$\half$) whereas a physical right-handed, helicity $j$ particle can
only correspond to the representation $[n,j+n]$.
The generalization of the statement to higher even dimensions will be stated
as the following corollary to the main theorem:
\smallskip
\noindent
{\bf Helicity-Chirality Correlation:} \it A physical field of the full group
corresponds to a highest weight state of the little group (given trivial
action
of translations), \rm and \it the eigenvalue of each generator common to the
full group and the little group remains unchanged under the projection of a
representation of the full group to a representation of the little group. \rm
\smallskip
\noindent
\bf Proof: \rm
Since the little group with trivial translations is just the orthogonal group
in two lower dimensions, a subset of the linearly independent raising
operators of the full group is exactly the complete set of linearly
independent raising operators of the little group, with the caveat that they
have to be appropriately Wick rotated to obtain the compact form.  For
$SO(2n+1,1)$ ($n >1$)
 as the full group, the subset of \ras\ without the last raising operator,
$L_{2n-1} + i L_{2n}$, upon Wick rotation, is the complete set of linearly
independent raising operators for $SO(2n)$.  For example, for $SO(5,1)$ the
raising operators are $L^+_1 \pm iL^+_2, L^+_3 + i L^+_4$, whereas for the
little group of trivial translations $\sim SO(4)$ they are $L^+_1 \pm L^+_2$.
Since the full set of raising operators annihilates the full group highest
weight, \it and \rm the projection from the full group to the little group is
an orthogonal projection (since the boost generator is the only non-common
generator and it commutes with the Cartan generators of the little group), it
follows that the little group state is a highest weight state of the little
group, and most importantly, the eigenvalue of the Cartan generators of the
little group is invariant under the projection (hence the consequences of the
corollary are most explicit in the weight notation).
To recapitulate, the helicity-chirality correlation in higher dimensions is
nothing but the fact that under projection of the field representation from
the full Lorentz group to a little group state, the eigenvalue of the Cartan
generators remains unchanged.

\medskip\noindent
In conclusion, we will demonstrate that our main theorem and corollaries
reproduce some familiar results of four dimensions:
\smallskip\noindent
\bt In weight notation, our result on the generalized helicity-chirality
correlation shows why a chiral left-handed  field transforming as
$(-\half,\half)_W$ ($[\half,0]$ in conventional $SU(2) \times SU(2) $ notation)
 corresponds to a helicity $-\half$ particle
 and a chiral right-handed field transforming as $(\half, \half)_W$
($[0,\half]$ in conventional notation)
corresponds to a helicity $ \half$ particle.
\smallskip\noindent
\bt Using $\half ({\bf J} \pm i {\bf K}) = {\bf A}, {\bf B}$ and the
definition $J_3 \Lambda \equiv J_{12} \Lambda = \lambda \Lambda$, with
$\lambda$ the helicity, we get, on using the physical field condition and
\ras\
\eqn\weinc{\eqalign{(L^+_1 - i L^+_2) \Lambda = 0 &\rightarrow (A_1 - i A_2)
\Lambda = 0 \rightarrow A_3 = -A \cr
(L^+_1 + i L^+_2) \Lambda = 0 &\rightarrow (B_1 + i B_2) \Lambda = 0
\rightarrow B_3 = B \cr}}
since ${\bf A}, {\bf B}$ generate independent $SU(2)$ algebras and $A_1 - i
A_2$ is the lowering operator for one and $B_1 + i B_2$ is the raising
operator
for the other.
Then, by definition $J_3 = A_3 + B_3 = B-A = \lambda$ which is Weinberg's
condition \WEIN\ .
\smallskip\noindent
\bt As mentioned earlier, the
translations are also generators of Abelian gauge transformations.  Requiring
them to be
trivial restricts us to gauge independent, finite dimensional
repesentations of the full group.  Our theorem and the corollary then tell us
what is the little group representation corresponding to this gauge
independent full group
representation.  For example, the representation corresponding to the field
strength tensor in the conventional ($SU(2) \times SU(2)$), Dynkin and weight
basis is:
\eqn\fieldstren{F^{\mu \nu}   = [1,0] + [0,1]  = (2,0)_D + (0,2)_D  = (-1,1)_W
+ (1,1)_W  }
which has $J_{12}$ eigenvalues $\pm 1$ and so is admissible as the $J_{12}$
eigenvalue remains
unchanged and corresponds to the correct helicity of the photon
 as we project to the little group state.
However, the vector potential corresponds to
\eqn\vecpot{A^\mu  = [\half,\half]  = (1,1)_D = (0,1)_W }
which has a $J_{12}$ eigenvalue of $0$ which does not correspond to a
transversely polarized helicity $\pm 1$ photon.

\newsec{Acknowledgments}

I would like to thank Professors H. Georgi, C. Vafa, A. Bohm,  V.
Hussin, M.A. del Olmo, L. Boya and the other organizers, and
the Graduate Student Council at Harvard University for facilitating my
participation in the colloquium.

\appendix{A}{ Weight Basis and Dynkin Basis}

In the Cartan basis for $SO(2n)$ labelled by the subscript $W$,
the Cartan subalgebra is generated by
$J_{12}, J_{34}, \ldots, J_{(2n-1,2n)}$ and a state is written by specifying
the ordered n-tuple of eigenvalues of the Cartan generators as
$(m_1,m_2,\ldots,m_n)_W$.  This basis is physical since the representation
specifies the effect of actual rotations.

The Dynkin basis labelled by the subscipt $D$ is useful due to Dynkin's
theorem which says that to every $n$-tuple of Dynkin indices that are
non-negative and integral, there corresponds a highest weight state of a
unique
irreducible representation (the converse is also true).
The Cartan matrix for the $SO(2n)$ groups is given in \SL :
\eqn\cartan{ A_{SO(2n)} =
\left(\matrix{ 2&-1&0&\ldots&0&0&0\cr
-1&2&-1&\ldots&0&0&0 \cr
0&-1&2&\ldots&0&0&0\cr
\ldots&\ldots&\ldots&\ldots&\ldots&\ldots&\ldots\cr
0&0&0&\ldots&2&-1&-1\cr
0&0&0&\ldots&-1&2&0\cr
0&0&0&\ldots&-1&0&2\cr   } \right), }
in which each row represents the eigenvalues of the positive simple roots
on commutation with the linearly independent diagonalizable
generators in the Dynkin basis.
Given a highest weight state in the Dynkin basis, the whole irrep. can be
constructed by subtracting rows of the Cartan matrix until all Dynkin
indices become non-positive.  This amounts to acting with the lowering
operators until the unique lowest weight state is obtained.

Using
\eqn\dy{a_i = 2  {{(\Lambda,\alpha_i)} \over {(\alpha_i,\alpha_i)}}}
where $\Lambda$ is the weight in the Weight basis, and $\alpha_i$ is the $i$th
positive simple root in the weight basis, we can get the Dynkin label for any
weight vector given in the weight basis.  Applying this to $SO(2n)$ we get
\eqn\dyw{\eqalign{a_1 &= m_n-m_{n-1} \cr
a_2 &= m_{n-1}- m_{n-2} \cr
&\vdots \cr
a_{n-1} &= m_2-m_1\cr
a_n & = m_2+m_1 \cr }}
For example, the $SO(3,1)$ spinor $(-\half,\half)_W$ is $(1,0)_D$ in Dynkin
notation and $(\half,\half)_W$ is $(0,1)_D$, and in general the
spinor corresponding to $+ \half$ eigenvalue under the Cartan generators, i.e.
$(\half,\half,\ldots,\half)_W$ is $(0,0,0\ldots,1)_D$ under the Dynkin
notation.

Inverting the equation, a state specified in the Dynkin basis can be written
in terms of the weight notation:
\eqn\wd{\eqalign{m_1& = \half(a_n-a_{n-1}) \cr
m_2&=a_{n-1} + m_1 \cr
m_3&=a_{n-2} +m _2 \cr
& \vdots \cr
m_n&= a_1 + m_{n-1} .\cr}}
For a highest weight, by Dynkin's theorem $a_i \geq 0$ so $m_n \geq m_{n-1}
\geq \ldots m_1$.

Finally, in four dimensions only (since $SO(4) \equiv SU(2) \times SU(2)$),
\dyw\ yields
 the translation between the Dynkin notation
and the conventional $SU(2) \times SU(2)$ notation (subscipted by $C$):
multiply each Dynkin index by
$\half$ i.e.$(0,1)_D$ of Dynkin notation becomes $[0,\half]_C$ which is a
right chirality, helicity
$\half$ object, and $(1,0)_D$ becomes $[\half,0]_C$ which is a left chirality,
helicity $- \half$
object.

\listrefs

\bye